\documentstyle[twocolumn,ifthen,floats,aps,prl,epsfig]{revtex}

\begin{document}
\draft

\newboolean{@epsimage}
\setboolean{@epsimage}{true}

\twocolumn[\hsize\textwidth\columnwidth\hsize\csname@twocolumnfalse\endcsname%

\title{Magnetically-induced reconstructions of the ground state in a few-electron
Si quantum dot}

\author{L.P. Rokhinson, L.J. Guo\cite{guoaddr}, S.Y. Chou and D.C. Tsui}

\address{Department of Electrical Engineering, Princeton University,
Princeton, NJ 08544}

\date{\today}

\maketitle

\begin{abstract}
We report unexpected fluctuations in the positions of Coulomb
blockade peaks at high magnetic fields in a small Si quantum dot. The
fluctuations have a distinctive saw-tooth pattern: as a function of
magnetic field, linear shifts of peak positions are compensated by
abrupt jumps in the opposite direction. The linear shifts have large
slopes, suggesting formation of the ground state with a non-zero
angular momentum. The value of the momentum is found to be well
defined, despite the absence of the rotational symmetry in the dot.
\end{abstract}


\vskip0pc]

The basics of Coulomb blockade (CB) phenomena can be understood
within the so-called "orthodox theory"\cite{meso-book91a}.  In
this theory, electron-electron interactions are hidden in the
charging energy and the electrostatic coupling is assumed to be
independent of0 the nature of the ground state, namely on the
particular distribution of electrons inside the dot. Early
experiments on large vertical quantum dots have already revealed
significant deviations from the "orthodox theory", noting that
several electrons can enter the dot at almost the same
energy\cite{ashoori92}. The deviations were later attributed to
localization of electrons in local minima of the confining
potential\cite{zhitenev97}. In a smooth confining potential,
magnetic field forces a redistribution of charges within the dot
to form quantum Hall edge states\cite{qd-edge}. Charge
redistribution is a focus of much theoretical work, especially in
the regime of high magnetic field\cite{highB-th}. A competition
between the attractive confining potential and the repulsive
electron-electron interactions is expected to produce a rich
variety of exotic patterns of charge distribution\cite{reimann99}.
In a recent experiment, high field (filling factor $\nu<1$)
redistribution of charges in a large vertical dot $>0.5$ $\mu$m
has been reported\cite{oosterkamp99}.

So far, most of the experiments were performed on two-dimensional
dots with weak and smooth confining potential electrostatically
created by gating. In such dots, the magnetic field $B$ dependence
of energy levels is dominated by orbital effects even at low $B$.
Recently, it has become possible to investigate quantum dots in a
different regime of very small size, strong confinement, and
strong local disorder\cite{leobandung95a}. This regime is realized
in three-dimensional Si dots with confining potential provided by
a sharp Si/SiO$_2$ interface. In these dots states with different
angular momenta are mixed due to the absence of rotational
symmetry, eliminating the linear in $B$ term of the orbital
energy, and the parabolic $B^2$ term is suppressed by strong
confinement. Thus, the $B$ dependence of the energy levels is
expected to be simple and to consist of linear Zeeman shifts.
Indeed, earlier we demonstrated that the shifts of CB peak
positions are dominated by the Zeeman effect\cite{rokhinson01a}.
In this work we focus on small fluctuations in the peak position,
which appears at high $B$ and low $T$ in a Si quantum dot with a
few electrons, $N\approx4-6$. The fluctuations have a distinct
saw-tooth pattern: as a function of $B$, fast linear shifts in the
peak position are interrupted by abrupt jumps in the opposite
direction. We show that the linear segments of the fluctuations
have orbital origin, indicating formation of states with a
relatively well defined non-zero angular momentum close to 1 or 2.
This is an unexpected result, taking into account the geometry of
the sample. Also, we show that the jumps are intrinsic to the dot,
presumably due to some magnetically induced rearrangement of
charges inside the dot.  We discuss the data within a simplified
single-particle picture. However, the observed phenomena is
clearly a many-body effect, and a proper treatment of
interactions, strong confinement and disorder is needed.

The small Si quantum dot is  lithographically defined from a
silicon-on-insulator wafer, and a sample layout is shown
schematically in the inset in Fig.~\ref{peak5}. A detailed
description of sample preparation can be found in
Ref.~\cite{leobandung95a}. The dot is three-dimensional,
asymmetric, and is elongated in the current flow direction. A
poly-Si gate is wrapped around the dot and is used to control the
number of electrons in the dot starting from $N=0$; the gate is
separated from the dot by 500\AA\ of SiO$_2$. The dot is connected
to two two-dimensional source and drain contacts via tunneling
barriers; the coupling is weak and even at the lowest temperature,
$T=60$ mK, the CB peaks remain thermally broadened.  From
excitation spectra we deduce single-particle energies to be $1-5$
meV, comparable to the charging energy $5-7$ meV.

The uniqueness of our dot is its small size ($\sim 100-200$ \AA\
long and $\lesssim100$ \AA\ in cross section) and the extremely
strong confining potential provided by the Si/Si0$_2$ interface
($\sim 3$ eV). The strength of the confinement is clearly seen
from the analysis of the first peak, see Fig.~2 of
Ref.~\cite{rokhinson01b}. The first energy level has a weak
parabolic $B$ dependence due to magnetic confinement
$(\hbar\omega_c)^2/E_0$, where $\omega_c=eB/m^*$ is the cyclotron
frequency. Characteristic energy $E_0$ depends on the direction of
$B$: $E_0\approx100$ meV for $B$ applied perpendicular to the
sample, $B_{\bot}$, and there is no measurable shift of the first
level for in-plane $B$ aligned with the current direction,
$B_{||}$. For both $B$ directions the magnetically induced
confinement is smaller than the Zeeman energy in the experimental
range of $B$.

\begin{figure}[tb]
\def\ffile{peak5}
\ifthenelse{\boolean{@epsimage}}{
\ifthenelse{\boolean{@twocolumn}}
{\epsfig{file=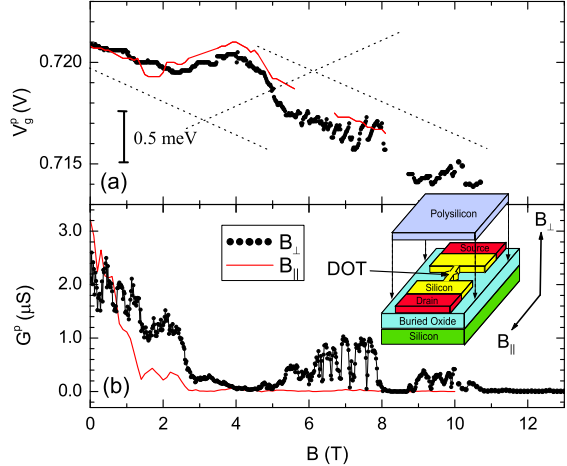,width=3.25in}}
{\vspace{0.7in}\centering\epsfig{file=\ffile.eps,height=4.5in}}}
{\ifthenelse{\boolean{@twocolumn}}
{\special{isoscale \ffile.wmf, 3.25in 2.6in}\vspace{2.5in}}
{\special{isoscale \ffile.wmf, 5.25in 6.5in}\vspace{6.5in}}}
\caption{(a) position and (b) amplitude of peak 5 plotted as a
function of magnetic field for two field directions: in-plane field
$B_{||}$ along the current direction (solid line) and $B_{\bot}$
perpendicular to the Si wafer (dots). Inset shows a schematic of the
sample. The data was measured at $T=60$ mK with 10 $\mu$V ac bias.
Dashed lines have slopes $\frac{1}{2}g^*\mu_B$, $g^*=2$.}
\label{\ffile}
\end{figure}

The number of electrons in the dot can be tuned between 0 and 30 with
the gate voltage $V_g$. The overall $B$ dependence of the energy
levels in this sample has been analyzed previously in
Ref.~\cite{rokhinson01a}. Evolution of the fifth CB peak with $B$
is plotted in Fig.~\ref{peak5}a. According to the previous analysis,
the $V^p_g(B)$ curve consists of 3 segments, each having
$\approx\frac{1}{2}g^*\mu_B/\alpha$ slope, where $g^*=2$ is the
$g$-factor in Si, $\mu_B=e\hbar/2m_0$ is the Bohr magneton and
$\alpha=14$ mV/meV. The two kinks at $B\approx2.5$ T and $B\approx4$
T have been identified as crossings of Zeeman split spin-up and
spin-down levels, originating from different single-particle energy
states.  Indeed, average peak position does not depend on the
direction of $B$. However, there is a noticeable difference between
the $V^p_g(B_{||})$ and $V^p_g(B_{\bot})$ curves at high $B$: $V^p_g$
changes linearly with $B_{||}$ but fluctuates as a function of
$B_{\bot}$. The appearance of these fluctuations is accompanied by a
dramatic increase of the peak amplitude $G^p$ by more than two orders
of magnitude, as shown in Fig.~\ref{peak5}b. The fluctuations are
most pronounced at our base temperature of 60 mK and are still
observable at 200 mK, while at $T\approx1$ K they are washed out
completely. No fluctuations were observed for $N>20$ or $N<4$. The
fluctuations are highly reproducible: the measured $V^p_g$ and $G^p$
from different scans are identical if the scans are performed within
the same cooldown; patterns from different cooldowns are similar,
although positions of the jumps may vary.

\begin{figure}[tb]
\def\ffile{jumps}
\ifthenelse{\boolean{@epsimage}}{
\ifthenelse{\boolean{@twocolumn}}
{\epsfig{file=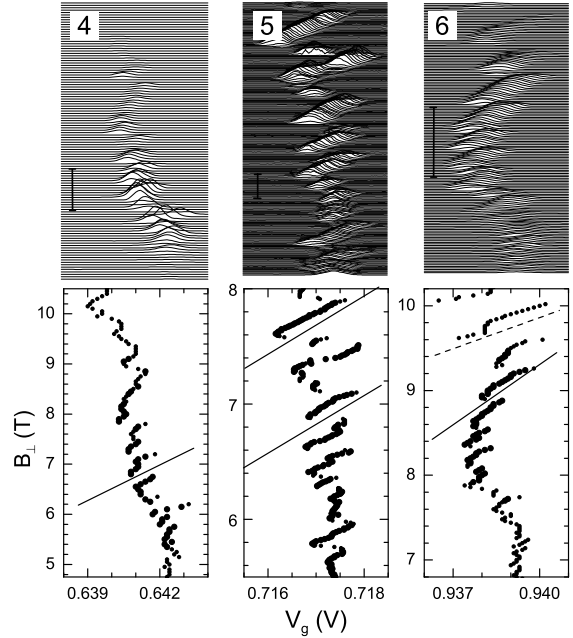,width=3.25in}}
{\centering\epsfig{file=\ffile.eps,width=4.5in}}}
{\ifthenelse{\boolean{@twocolumn}}
{\special{isoscale \ffile.wmf, 3.25in 3.5in}\vspace{3.4in}}
{\special{isoscale \ffile.wmf, 5.25in 4in}\vspace{4.5in}}}
\caption{Evolution of peaks 4, 5 and 6 as a function of
perpendicular magnetic field $B_{\bot}$.  On the top panels, a
series of $G$ vs. $V_g$ curves are plotted, each linearly offset
with $B_{\bot}$. The bars are 4 $\mu$S scales. In the bottom
panels, peak positions are extracted and plotted as a function of
$B_{\bot}$ for the same range of $B_\bot$ as in the corresponding
top panels. Solid and dashed lines have the slopes
$\frac{1}{2}\hbar\omega_c/B$
 and $\hbar\omega_c/B$.}
\label{\ffile}
\end{figure}

Fluctuations of $V_g^p$ for peaks 4, 5 and 6 are magnified in
Fig.~\ref{jumps}. The fluctuations have a distinctive saw-tooth
pattern: $V_g^p$ increases linearly with $B_{\bot}$, then
decreases abruptly. Most of the linear segments have slopes close
to 4 mV/T (0.3 meV/T) or 8 mV/T (0.6 meV/T). These slopes are much
larger than $\frac{1}{2}g^*\mu_B=0.06$ meV/T and are grouped
within 15\%\  around $\frac{1}{2}\hbar\omega_c/B$ and
$\hbar\omega_c/B$. The linear shifts are interrupted by abrupt
jumps of the peak position, which occur within $<20$ mT. These
jumps compensate the linear shifts and, on average, the peaks
follow the weak $V_g^p$ vs $B_{||}$ field dependence due to the
Zeeman shift of energy levels.

\begin{figure}[tb]
\def\ffile{rotBz}
\ifthenelse{\boolean{@epsimage}}{
\ifthenelse{\boolean{@twocolumn}}
{\epsfig{file=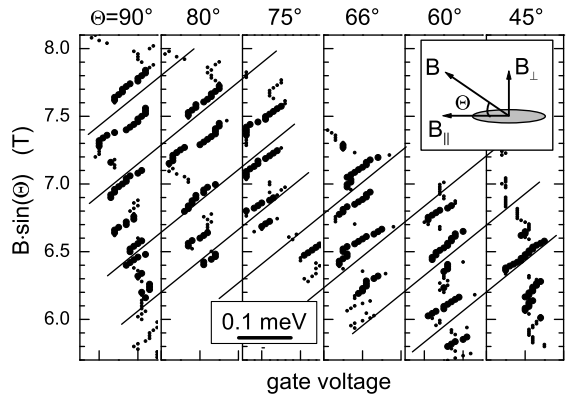,width=3.25in}}
{\centering\epsfig{file=\ffile.eps,width=5.0in}}}
{\ifthenelse{\boolean{@twocolumn}}
{\special{isoscale \ffile.wmf, 3.25in 2.5in}\vspace{2.4in}}
{\special{isoscale \ffile.wmf, 5.25in 5.5in}\vspace{5.5in}}}
\caption{Position of peak 5 as a function of
$B_{\bot}=B\cdot\sin\Theta$ plotted for different angles $\Theta$.
The angle is defined in the inset. The lines have a slope
$\frac{1}{2}\hbar\omega_c/B$ and are guide to the eye.}
\label{\ffile}
\end{figure}

Orbital effects can be distinguished from spin effects by studying
the evolution of peak positions in a tilted magnetic field.  In
Fig.~\ref{rotBz}, $V_g^p$ of peak 5 is plotted as a function of
$B\sin(\Theta)$ for different angles $\Theta$ between $B$ and $I$.
Linear peak shifts depend on the perpendicular component
$B_{\bot}\approx B \sin(\Theta)$, rather than on the total filed
$B$ (solid lines provide a guide for the eye). Curves at different
angles do not scale with $\sin(\Theta)$ exactly, as is expected
for a three-dimensional structure. Taking into account that the
energy shift due to the magnetic confinement is negligible, we
attribute the slopes $\approx\frac{1}{2}\hbar\omega_c/B$ to the
formation of states with a non-zero angular momentum $m\approx1$.
Segments with larger slope $\approx\hbar\omega_c/B$ have also been
observed and we attribute them to the formation of states with
$m\approx2$.

Fast linear shifts of $V^p_g$ as a function of $B_{\bot}$ are
interrupted by abrupt jumps in the opposite direction.  These jumps
occur within $\Delta B<20$ mT and, if attributed to a state with
large angular momentum, would correspond to an unrealistic $m>10$.
Ground state (GS) energy of the dot should be a continuous function
of all variables, including $B$, unless the GS is bi-stable.
Bi-stability of the GS should reveal itself through a hysteresis of
the conductance in $V_g$ or $B$ scans. Experimentally, the traces
were found to be identical, independent of the scan direction. Thus,
jumps in the peak position should reflect either abrupt changes in
the environment or abrupt changes in the electrostatic coupling
between the dot and the environment.

In the CB regime, conductance is non-zero only when the
electrochemical potential of the dot equals the electrochemical
potential of the leads. Peak position $V_g^p$ is determined from the
condition\cite{beenakker91}
\[
\frac{e [(N-1/2)-C_gV_g^p]}{C_{\Sigma}} +
[E(N)-E(N-1)]+e\phi_{ex}=E_F,
\]%
where $e^2/2C_{\Sigma}$ is the Coulomb energy for one additional
electron, $C_g$ and $C_{\Sigma}$ denote respectively the gate and
the total capacitances, $E(N)$ is the kinetic energy of the state
with $N$ electrons, $E_F$ is the Fermi energy in the leads, and
$\phi_{ex}$ is a potential induced by external charges. According
to the above equation, there are three possibilities for a jump of
$V_g^p$: i) an abrupt change in $E_F$, ii) a discrete change of
the background charge distribution and, thus, $\phi_{ex}$, or iii)
an abrupt change in the electrostatic coupling $C_g$. It is easy
to rule out $E_F$ as a source of the jumps, since depopulation of
Landau levels should lead to the jumps of $V_g^p$ in the opposite
direction.

\begin{figure}[tb]
\def\ffile{pk5}
\ifthenelse{\boolean{@epsimage}}{
\ifthenelse{\boolean{@twocolumn}}
{\epsfig{file=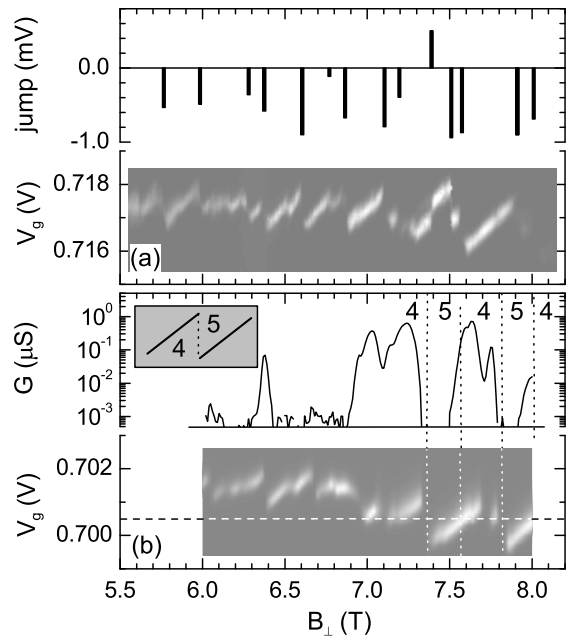,width=3.25in}}
{\centering\epsfig{file=\ffile.eps,width=4.5in}}}
{\ifthenelse{\boolean{@twocolumn}}
{\special{isoscale \ffile.wmf, 3.25in 3.4in}\vspace{3.3in}}
{\special{isoscale \ffile.wmf, 5.25in 4.5in}\vspace{4.5in}}}
\caption{Gray-scale plots show conductance as a function of $V_g$ and
$B_{\bot}$ near peak 5.  The data was taken by scanning $V_g$ at
different fixed values of $B_{\bot}$.  In (a) the amplitude of each
$V_g^p$ jump is extracted.  In (b) conductance as a
function of $B_{\bot}$ was measured at a fixed
$V_g=0.7005$ V, marked by the dashed horizontal line in the
gray-scale plot. Note the logarithmic scale of $G$. The vertical
dotted lines separate ground states with different number of
electrons along the curve. The data in (a) and (b) were collected
during different cooldowns.}
\label{pk5}
\end{figure}

It is appealing to attribute the jumps to a magnetically induced
depopulation of a charge trap. However, such an explanation is
inconsistent with the data. One has to assume the existence of a
large two-dimensional trap capacitively coupled to the dot, which
does not participate in the charge transport. Qualitatively, there
are rather general arguments against such a scenario. If jumps are
related to the magnetic confinement of electrons in the trap, we
expect the frequency of the jumps to increase with $B$, and there
should be no low-field cut-off for their appearance. Experimentally,
the jumps appear suddenly at $B\approx5$ T and the frequency of the
jumps slightly decreases with $B$; there are no jumps at lower
fields. In Si nanostructures, traps are formed inside the oxide
layer\cite{double-dot}, and it is hard to explain
their existence only at high $B_{\bot}>5$ T and in a limited range of
gate voltages $0.5$ V $<V_g<1$ V. Quantitatively, in order to account
for the data, one has to assume the existence of a) several traps, b)
all within the tunneling range from the dot, inconsistent with the
geometry of the sample.  Electrostatically, discharging of a trap
leads to a fixed shift $\Delta V_g^p=-C_c/C^{ex}_{\Sigma}\cdot
e/C_g$, where $C_c$ is the dot--trap cross capacitance, and
$C_{\Sigma}$ and $C^{ex}_{\Sigma}$ are the corresponding total
capacitances.  In Fig.~\ref{pk5}a, values of the $V_g^p$ shifts are
extracted for each jump in the range 5 T $<B_{\bot}<$ 8.2 T.  First,
there is a {\it positive} shift at $B_{\bot}=7.4$ T, inconsistent
with the trap discharging model. Second, amplitude of the negative
shifts ranges from 0.2 to 1.0 mV with no systematic pattern. Thus, a
single trap cannot account for the observed oscillations. Subsequent
discharging of several traps can be ruled out using the following
arguments. In the $V_g-B$ plane the number of electrons in the dot
differs by one across the jump, as shown schematically in the inset
in Fig.~\ref{pk5}b. Following the analysis of a two-dot
system\cite{hofmann95}, there should be a peak in the conductance
along the jump, unless the trap and the dot are connected by
tunneling (in this case an electron tunnels between the dot and the
trap and the total number of the electrons in the dot-trap system
remains constant). Experimentally, there is no peak in conductance if
$B$ is swept across the jump at a fixed $V_g$, as shown in
Fig.~\ref{pk5}b.  An assumption that several large traps are in a
close proximity of the dot is inconsistent with the small size of the
device.

It is very suggestive that all the observed effects, namely jumps
in peak position, enhanced conductivity, appearance of large
slopes and an apparent mutual compensation of shifts and jumps,
are related and originate from the dot. Presumably, the jumps are
related to the magnetically induced rearrangement of electron
density inside the dot. In our dots, we expect strong fluctuations
of the confining potential, because Si/SiO$_2$ interface roughness
directly translates into large fluctuations of the local
potential. At low $B$ fields the electron wavefunction is spread
over the entire dot. At high $B$ fields the extent of the
wavefunction is determined by the magnetic length
$l_m=(\hbar/eB)^{1/2}$. When $l_m$ becomes smaller than the
average distance between electrons, Coulomb repulsion favors
localization of electrons. We expect fluctuations of the local
potential to facilitate the redistribution of charge. Strong
increase in the peak height hints that electrons are rearranged
closer to the dot boundary, where their wavefunctions have larger
overlap with electrons in the leads. It has been pointed out in
Ref.~\onlinecite{tans98} that electrostatic coupling to the gate
$C_g$ is not necessarily a constant, but depends on the particular
distribution of electron density inside the dot. In order for a
peak to shift by 1 mV, only a small change $\Delta
C_g/C_g\approx\Delta V_g^p/V_g^p<0.2\%$ is required.

The most unexpected and surprising result is the appearance of
states with non-zero angular momentum. In the absence of
rotational symmetry, states with different angular momenta are
mixed, and angular momentum is not a good quantum number. However,
the dot asymmetry becomes less important if an electron can
complete a classical cyclotron orbit faster than the time
$\tau=L/\sqrt{2E/m^*}$ needed to traverse the dot. For an electron
with kinetic energy $E=4$ meV in a dot of size $L=200$ \AA, the
cross over $\omega_c^{-1}<\tau$ should occur at $B\approx5$ T.
Note that $E\approx N \Delta$ increases with the number of
electrons (here $\Delta$ is the level spacing), leading to the
increase of the cross over field with $N$. We speculate that, for
a few electrons in the dot, high $B$ provides a mechanism to
suppresses the mixing of different angular momentum states.

Electron-electron interactions in our sample are rather strong
($e^2/4\pi\epsilon\epsilon_0r=12$ meV for $r=100$ \AA). They reveal
themselves in the spontaneous polarization of the $N=6$ ground state
at low fields and in the suppression of the corresponding CB peak due
to spin blockade\cite{rokhinson01a}. Thus, a many-body description of
the ground state is required.

To summarize, we explored a quantum dot in a new regime of small
size, strong confinement and strong electron-electron
interactions. We observed fluctuations in the CB peak positions at
high $B$ for a few-electron states. The fine structure consists of
fast linear shifts, followed by abrupt jumps in peak positions. We
argue that the fluctuations are intrinsic to the dot and reflect
changes in the many-body wavefunction.  We attribute jumps to the
magnetically induced spatial rearrangement of charges inside the
dot. The linear segments have large slopes, reflecting unexpected
formation of states with relatively well defined non-zero angular
momentum. The observed phenomena is a many-body effect and is
clearly  beyond the description of a single-particle picture.

The authors are grateful to Boris Altshuler for critical reading of
the manuscript and valuable comments. The work was supported by the
ARO, ONR and DARPA.


\end{document}